\documentstyle[prl,aps,psfig,floats,twocolumn]{revtex}
\begin{document}
\title{Cooper pair delocalization in disordered media}
\author{M.A.N. Ara\'ujo $^{1}$, M. Dzierzawa $^{2}$, 
C. Aebischer $^{3}$, and D. Baeriswyl $^{3}$}
\address{
$^{1}$Departamento de F\'{\i}sica da Universidade de \'Evora, 7000  
\'Evora, Portugal \\
$^{2}$Institut f\"ur Physik, Universit\"at Augsburg, D-86135 Augsburg, 
Germany \\
$^{3}$Institut de Physique Th\'eorique, Universit\'e de Fribourg, 
P\'erolles, CH-1700 Fribourg, Switzerland
}
\maketitle
\begin{abstract}
We discuss the effect of
 disorder on the coherent propagation of the bound state of two attracting 
particles. 
It is shown that a result analogous to the Anderson theorem for dirty 
superconductors 
is also valid for the Cooper problem, namely, that the pair wave function 
is extended beyond the single-particle localization length if the latter 
is large.
A physical justification is given in terms of the Thouless block-scaling 
picture of localization. These arguments are supplemented by numerical 
simulations.
With increasing disorder we find a transition from a regime in which the 
interaction delocalizes the pair to a regime in which the interaction 
enhances localization. 
\end{abstract}

\section{Introduction}

It has long been known that the wave function of an electron moving in a  
random potential 
 becomes spatially localized. This effect was first predicted by Anderson 
\cite{anderson}  
and is termed {\it Anderson localization}. In one and two dimensions {\it 
all} quantum 
states are localized 
in the presence of {\it any} amount of disorder while in three dimensions 
localization occurs only 
above some critical disorder. 
While this phenomenon is now fairly well understood in a single-particle 
picture, the inclusion of 
 interactions in  the  disordered many-body system  is a non-trivial 
problem. Insight can be gained  by studying  the simpler case of just two 
interacting particles in a random 
potential. 
In this context, it has  recently been claimed that the interaction can 
actually  lead to a
 {\it delocalization} effect, in the sense that the spatial extent of the 
two-body wave function is 
larger than the single-particle localization length \cite{shep}. This 
delocalization was found  for both attractive and 
repulsive interactions, at least for some of the eigenstates in the
 continuous spectrum of 
energy eigenvalues.  
 Some authors objected to this finding \cite{romer}
 and 
this point  is still being debated. Furthermore, the relevance of these 
results for  the many-body problem
has not  yet been clarified.

In this paper our main concern is the case  of an {\it attractive} 
interaction. The concept 
of  a two-particle bound state  (the Cooper pair) plays an important role 
in the theory of superconductivity. 
There is theoretical and experimental evidence for the existence of 
superconductor-insulator transitions, where localized states combine 
coherently into a superconducting condensate with a finite superfluid 
density. This finding motivates the question addressed in this paper: 
will the bound state of two attracting particles be extended over 
distances larger than the single-particle localization length? We find 
that a result analogous to the Anderson theorem for localized 
superconductors is valid for this problem in the limit of large 
single-particle localization lengths. 
By increasing the disorder we find a transition from a regime in which 
the interaction {\it increases} the localization length to a regime in 
which the interaction {\it reduces} the localization length.

\section{Localized superconductors}

The Hamiltonian for a disordered conventional ($s$-wave) superconductor is
\begin{equation}
H=H_{0} - V \int \psi_{\downarrow}^{\dagger}({\bf r}) 
\psi_{\uparrow}^{\dagger}({\bf r})
\psi_{\uparrow}({\bf r})\psi_{\downarrow}({\bf r}) d{\bf r} 
\label{bcs}
\end{equation} 
where $H_{0}$ represents the single-electron part including a spatially 
random   external potential and the interaction parameter $V$ is taken as 
positive. The attractive interaction could 
 be  due to  exchange of phonons  or purely electronic mechanisms
but its origin does not concern us here. Depending on the disorder,
the eigenstates $\phi_{n}({\bf r})$ of $H_{0}$ can be
extended  or localized (with localization length $\rm  L_{c}$ at the 
Fermi level).

At T=0 the system described by (\ref{bcs})
is a superconductor with spatially constant order parameter $\Delta ({\bf 
r}) = V \\
< \psi_{\uparrow}({\bf r}) \psi_{\downarrow}({\bf r})> = \Delta$ 
when the condition 
\begin{equation}
\rm
\Delta N_{F} L_{c}^d \gg 1
\label{valanderson}
\end{equation} 
holds \cite{lee}. $\Delta$ is then given by the same expression as that 
for a clean superconductor. 
This is the Anderson theorem. The important message of equation 
(\ref{valanderson}) is that  if the disorder is strong enough to localize 
the single-particle states,  the superconducting
order parameter $\Delta$ and critical temperature $\rm T_{c}$ will 
remain  unaffected as long as
 the number
of single-particle states in the energy range $\Delta$ contained in  the 
localization
volume $\rm L_{c}^d$ is still
large. (The 
superfluid density, on the other hand, is greatly reduced.)
If the amount of disorder is further increased so that 
(\ref{valanderson}) is no longer valid
  the
  superconducting order parameter will fluctuate strongly in space and the 
critical temperature will
be lowered. 
 
Note that because $\Delta({\bf r}) \propto <\psi_{\uparrow}({\bf 
r})\psi_{\downarrow}({\bf r})>$ is the 
wave function of the condensate, the Anderson theorem tells us that the 
attractive interaction is
delocalizing
the Cooper pairs. Keep in mind, however, that in this derivation the 
Cooper pairs are strongly
 interacting (and 
actually overlapping) with each other,
 forming
a correlated li\-quid. In BCS theory the concept of  Cooper pair only has 
the formal significance that 
 strong pair correlations exist between the particles in  phase space. 
In the regime of validity of the Anderson theorem 
  $\Delta$ is given by
\begin{equation}
1 = V \sum_{n} \frac{1}{2 \sqrt{\Delta^2 + \xi_{n}^2}} = V \sum_{n} 
\frac{1}{2 \epsilon_{n}}.
\label{delta}
\end{equation} 
The denominator of this expression is the quasi-particle excitation 
energy. The
binding energy of an ''isolated`` Cooper pair is given by an expression of 
the same form as this one 
(see eq.(\ref{cond}) below)
but with a 
different denominator. This is because equation (\ref{delta}) takes into 
account the interactions between 
Cooper pairs.

Experimentally, it has been found that superconductivity persists up to 
the  Anderson
metal-insulator  transition
\cite{um,dois,tres,quatro}. These systems  exhibit activated conductivity 
above $\rm T_{c}$.
The coexistence of  superconductivity  and localization has also been 
observed in underdoped
high-$\rm T_{c}$ superconductors \cite{boe,bes,ando,fukuzumi}. 

\section{Cooper problem in a random potential}

In what follows
the problem of two interacting electrons
 in a random potential will be addressed for the specific case of  an 
attractive interaction.
  The aim is to see  how the interplay of 
 disorder and interaction affects the coherent propagation of the  
electrons in the ground state. 
The attractive interaction will be assumed  to be short-ranged. We can 
describe this system by an
Anderson-Hubbard
 Hamiltonian
\begin{equation} 
H =  -t\sum_{<{\bf r},{\bf r}'>,\sigma} c_{{\bf r},\sigma}^{\dagger} 
c_{{\bf r}',\sigma} - U \sum_{{\bf r}}
 n_{{\bf r},\uparrow} n_{{\bf r},\downarrow}
+ \sum_{{\bf r},\sigma} \epsilon_{{\bf r}} n_{{\bf r},\sigma}
\label{ham}
\end{equation}
with $\epsilon_{{\bf r}}$ representing the site energies randomly 
distributed over a width $W$, and $U>0$.
The single-particle eigenfunctions $\phi_{n}$ are 
assumed to be localized by the 
disorder with a localization length
$\rm L_{c}$.

 We search for the  two-particle
ground state
\begin{equation}
\hat{H} \Psi({\bf r}_{1},{\bf r}_{2}) = E_{0} \Psi({\bf r}_{1},{\bf r}_{2})
\end{equation}
where ${\bf r}_{1}$ and ${\bf r}_{2}$ are the coordinates of the 
electrons. Because the bound state
is a spin singlet, $\Psi({\bf r}_{1},{\bf r}_{2})$ is a symmetric 
function of
its arguments and  has the general form
\begin{equation}
\Psi({\bf r}_{1},{\bf r}_{2}) = \sum_{n,m} g_{n,m} \phi_{n}({\bf 
r}_{1})\phi_{m}^{*}({\bf r}_{2})
\label{psi}
\end{equation}
where $\phi_{n}$ are normalized single-particle eigenfunctions of 
(\ref{ham})
 with energy eigenvalues $\xi_{n}$.
 In order to simulate the presence of a Fermi sea, it is
assumed that the two electrons can only be paired in states $\phi_{n}$ 
which lie above the Fermi surface.
According to the values of the parameters in (\ref{ham}) we recognize 
several different regimes. If 
$U \gg t$ (strong attraction) then the electrons are tightly bound and 
move together like a heavy boson
with hopping amplitude $t^2/U$ in an environment  with disorder W.  This 
boson would then become easily
 localized.  In the remainder of the paper we concentrate on the regime 
in which  
 the interaction is not strong ($U \leq t$). In this case 
 $\Psi$ is essentially the result of electron pairing in 
time-reversed single-particle eigenstates of (\ref{ham}) for not too 
strong disorder. This can be seen as follows.

The Schr\"odinger equation for the Cooper pair wave function is 

\begin{eqnarray}
\nonumber
& \Psi({\bf r}_{1},{\bf r}_{2}) &  =  \\ 
&  -\sum_{n,m}  \frac{ U \sum_{\bf r'}   
\phi_{n}^{*}({\bf r'}) \phi_{m}({\bf r'}) 
 \Psi({\bf r'},{\bf r'})}{ E_{0} - \xi_{n}- \xi_{m}} \phi_{n}({\bf 
r}_{1})\phi_{m}^{*}({\bf r}_{2}). 
\label{sch}
\end{eqnarray}

It admits  the  solution $\Psi({\bf r},{\bf r}) = \rm 
const.$ if the condition 
\begin{eqnarray}
1 & = & -U \sum_{n} \frac{ \mid \phi_{n}({\bf r} \mid^2 }
{ E_{0} - 2 \xi_{n}} \nonumber \\
& = & -U \int d\xi \frac{ N(\xi,{\bf r}) }
{ E_{0} - 2 \xi}
\label{cond}
\end{eqnarray}
holds. (Here $N(\xi,{\bf r})$ denotes the local density of states at the 
point ${\bf r}$). Thus a result analogous to the Anderson theorem is
valid for the function $\Psi({\bf r},{\bf r})$.
In such a case
 the solution to (\ref{sch}) is 
\begin{equation}
\Psi({\bf r}_{1},{\bf r}_{2}) \propto \sum_{n} 
\frac{\phi_{n}({\bf r}_{1}) \phi_{n}^{*}({\bf r}_{2})}
{E_{0}-2\xi_{n}}
\label{extended}
\end{equation}
and $\Psi$ has no overlap with the state $\phi_{n}({\bf r}_{1}) 
\phi_{m}^{*}({\bf r}_{2})$ if $n \neq m$. The condition (\ref{cond}) can 
be satisfied if the integral of the local density of states $\int d \xi 
N(\xi,{\bf r})$
over an energy interval of order $\mid E_{0}\mid $ does not depend on ${\bf r}$. 
This is possible  
even if the disorder is strong enough to localize the single-particle 
states  as long as the condition
 
$$
\mid E_{0}\mid {\rm N_{F} L_{c}^d} \gg 1
$$
holds. The binding energy $\mid E_{0}\mid $ will then be the same as 
that for a 
clean system. So we reach the conclusion that the attractive interaction 
can delocalize the pair or, at least, increase its localization length.

\section{Delocalization by interactions}

The delocalization by interactions due to the attractive interaction can 
be understood within the block-scaling picture of localization introduced 
by Thouless \cite{thou}.
In a recent paper \cite{imry}, Imry has used the block-scaling 
picture in order to argue that interactions (attractive or repulsive) 
should, in some cases, delocalize some of the eigenstates of the 
continuous spectrum
of the pair of electrons. In what follows we extend the argument to the 
case of a Cooper pair.

 Suppose the 
system is divided into blocks of linear  size $\rm L_{c}$ (measured in 
units of the lattice spacing) so 
that  the mean level spacing in a block
is $\Delta_{1}=B/L_{c}^{d}$ with $B$ equal to the band width. We can then 
solve the Cooper problem, as above,
 for
each block by pairing  the two electrons in  time-reversed states  
localized inside the  
block. If the fluctuation of the binding energy is smaller than 
 the effective scattering amplitude of the pair between  
blocks then the pair will be extended over many blocks. 
We denote the   Cooper pair in the $j$-th block by  $\Psi_{j}$. Next, we 
estimate the scattering
 amplitude $t_{eff}$, {\it due to the interaction}, 
of the Cooper pair between two neighbouring blocks as
\begin{eqnarray}
t_{eff}  & = & -<\Psi_{j} \mid U \sum_{{\bf r}} n_{{\bf r},\uparrow} 
n_{{\bf r},\downarrow}
\mid \Psi_{j'}> \nonumber \\
 & = & 
-U \sum_{n,n'} g_{nn} g_{n'n'}
 \sum_{{\bf r}} \mid \phi_n({\bf r}) \mid ^2 \mid \phi_{n'}({\bf r}) \mid ^2
 \nonumber \\
& \approx &  - \frac{U}{\rm  L_{c}^d} \sum_{n,n'} g_{nn} 
g_{n'n'}.
\label{hopp}  
\end{eqnarray}
We now note that $g_{nn}$ depends smoothly on  $n$ and has
  no nodes because it is the wave function of a ground state, namely
$$
g_{nn} \propto \frac{1}{E_{0} - 2 \xi_{n}}.
$$
The normalization
of $\Psi_{j}$ implies
 $g_{nn} \sim (\mid E_{0} \mid / \Delta_{1})^{\frac{1}{2}}$ and
 the number of terms in   the sum  in (\ref{hopp}) is large, of the order of
 $(\mid E_{0} \mid / \Delta_{1})^2$. All those
terms interfere constructively yielding
$$
t_{eff} \sim 
 \frac{\mid E_{0}\mid }{\Delta_{1}} \frac{U}{\rm  L_{c}^d} \sim \frac{U 
\mid E_{0}\mid }{B}. 
$$
The fluctuation of the binding energy $\mid E_{0}\mid $ due to the randomness 
in $\xi_{n}$ is of the order of
$\delta E \sim \sqrt{\mid E_{0}\mid \Delta_{1}}$ (see the Appendix). 
Thus the condition for delocalization is obtained as 
$$
\frac{t_{eff}}{\delta E} \sim \sqrt{\frac{\mid E_{0}\mid }{\Delta_{1}}} \ 
\frac{U}{B} \gg 1.
$$
The ratio $U/B$ is the BCS product $\rm N_{F} V$.
 The following  points should be noted:

 $(i)$ 
There are no  phase correlations to consider 
in the sum (\ref{hopp}) because all the terms are real and negative;

 $(ii)$  The large
  effective hopping amplitude $t_{eff}$ resulted from 
 $g_{nn}$ being a smooth function of $n$ with no nodes. In other words,
 {\it because  we have been  considering the  ground state}. 

$(iii)$ If the sign of the interaction $U$  is reversed,to make it 
repulsive,  we do not expect
any delocalization to occur in the ground state.
   The reason for this is that 
$\Psi_{j}$ for each  block (using time-reversed pairing) would now 
essentially involve only the
  single-particle eigenstate $\phi_{0}$
with the lowest energy:
 $\Psi_{j} ({\bf r}_{1},{\bf r}_{2})
\approx \phi_{0}({\bf r}_{1})\phi_{0}^{*}({\bf r}_{2})$. Then $t_{eff} 
\sim U/{\rm  L_{c}^d} \sim
\Delta_{1}$ if $U \approx B$. If we remove the constraint of 
time-reversed pairing then Imry's
argument would predict delocalization only if the pair has a certain 
excitation energy above the 
Fermi level \cite{imry,jaquod}. So it would not be in the ground state.

\section{Numerical results}

We have also performed numerical calculations of the Cooper pair wave function
for 1D systems of up to $L = 100$ sites using the Lanczos algorithm. In 
order to impose the constraint of a filled Fermi sea the single-particle
eigenstates for a given realization of the disorder potential
are calculated and the eigenenergies of the states below the Fermi energy
are shifted by a large amount. Thus only the one-particle states
above the Fermi surface are accessible for the Cooper pair.
Before doing the Lanczos diagonalization of the two-particle problem
the Hamiltonian is transformed back
to real space representation where the number of nonzero matrix
elements is much smaller than in the basis of the eigenstates of the
non-interacting disordered system. 

In order to determine the spatial extent of the Cooper pair wave function
with respect to both the relative and the center-of-mass coordinates 
we have calculated two different quantities. 
The first one is the 
participation ratio 
$$
r=\frac{1}{\sum_{\bf r} \mid \Psi({\bf r},{\bf r})\mid^4}
$$
where $\Psi({\bf r},{\bf r})$ is normalized to unity. Since only the diagonal
part $({\bf r_1}={\bf r_2})$ of the wave function is involved in the 
calculation the participation ratio can also be interpreted as the
localization length of the Cooper pair.

The second quantity, which is related to the relative coordinate of the 
two electrons,
is the average size $d$ of the Cooper pair defined as
$$
d=\sum_{{\bf r}_{1},{\bf r}_{2}} \mid 
\Psi({\bf r}_{1},{\bf r}_{2}) \mid^2 \mid {\bf r}_{1}-{\bf r}_{2}\mid.
$$

Our results are obtained for a system of 100 sites and each data
point represents the average over 200 realizations of the disorder
potential. Since the fluctuations of the participation ratio $r$ are 
very large we found it more convenient to average $\log r$
instead of $r$ itself. 

Fig. 1 shows the behavior of $\log r$ as a function of $U / t$ for
different values of the disorder strength $W / t$. 
While in the case of strong disorder the interaction leads to a 
decrease of the participation ratio, for smaller values of $W / t$ an
enhancement of $\log r$ is observed, at least for not too large $U / t$.
This is in qualitative agreement with our analytical arguments.
One should however keep in mind that the calculations are done for a
rather small system and that finite size effects 
can be of importance, especially in the weak disorder region where
the localization length becomes large.

In Fig. 2 we see that the size of the Cooper pair 
decreases rapidly with increasing interaction strength.
The values for $U / t \rightarrow 0$ which should be of the order of the
one-particle localization length are considerably reduced due to finite 
size effects.  

\begin{figure}
\centerline{\psfig{file=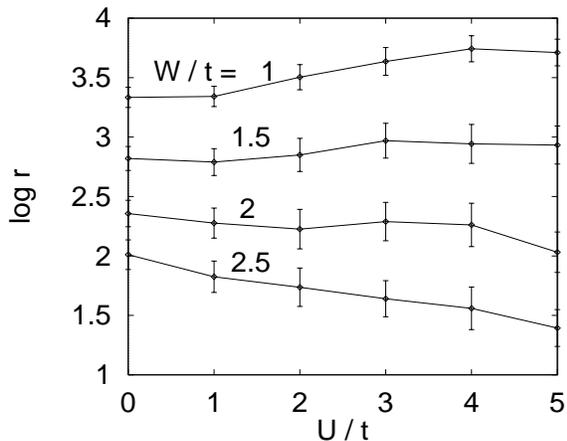,width=8.5cm,angle=0}}
\caption{Participation ratio $r$ of the diagonal component of the Cooper
pair wave function as a function of $U / t$ for several values
of the disorder strength $W / t$.}
\label{Fig.1}
\end{figure}

\begin{figure}
\centerline{\psfig{file=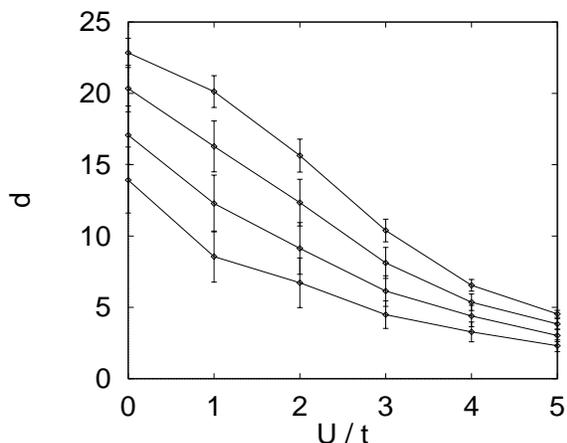,width=8.5cm,angle=0}}
\caption{Average size $d$ of the Cooper pair as a function of
$U / t$ for $W / t = 1, 1.5, 2, 2.5$ (from top to bottom).}
\label{Fig.2}
\end{figure}

\section{Conclusion}

The interplay of disorder and interaction is a complex problem and  
therefore it is useful to start with the 
two-body problem in a disordered medium. 
The effect of a repulsive  interaction seems to be very different from 
that of an attractive interaction
since the latter can induce propagation of the bound state of two 
particles (their ground state). 
The binding
energy and spatial extent of the  bound state 
is insensitive to disorder beyond the point where single-particle states 
become localized. This result
for the Cooper pair corresponds to
 the familiar Anderson theorem for the many-body problem of dirty 
superconductors and is valid
 in a regime 
where  the attraction and  the disorder are not too strong so that 
electrons are paired in
 time-reversed single-particle eigenstates. 
These localized states can then be combined coherently into an extended 
pair wave function (eq.(\ref{extended})). In analogy with the case of 
dirty superconductors discussed in \cite{lee} where the superfluid 
density is greatly reduced by the disorder, we also expect the energy of 
the wave function (\ref{extended}) to be much less sensitive to the 
boundary conditions than that in a clean system. In other words, disorder 
has a stronger effect on the sensitivity to boundary conditions than on 
the localization length.
It is possible to go away from this ''Anderson regime`` in
two ways.
If the interaction is increased then both the binding energy  and the 
effective mass of the 
pair increase, so the pair becomes more localized. On the other hand, if 
the interaction is kept not too 
strong but
the disorder is increased, the  binding energy and the localization 
length are reduced.
\vskip 5mm

{\large \bf Acknowledgements}
\vskip 5mm
M.A.N.A. would like to thank the hospitality of the Institut de physique
th\'eorique, Fribourg, Switzerland, where part of this work was carried 
out.
\vskip 5mm

{\large \bf Appendix}
\vskip 5mm

We want to prove that $\delta E \sim \sqrt{E \Delta_{1}}$.
The binding energy 
E is obtained from the equation
\begin{equation}
\frac{1}{U}  =  \sum_{n} \frac{1}{E +2 \xi_{n}}
 =  
\int \frac{N(\xi) d\xi}{E +2 \xi}. 
\end{equation}
The density of states is $N(\xi) =N_{0}(\xi) +n(\xi)$ where
$N_{0}(\xi)$ is the average density of states  and $n(\xi)$ the noise due
to disorder.
Then $E = E_{0}+\delta E$ where $E_{0}$ is the binding energy derived 
from $N_{0}(\xi)$.
Expanding the right-hand side of the above equation for small $
\delta E$ and taking into account the definition of $E_{0}$ we arrive at
\begin{equation}
\delta E \int \frac{N_{0}(\xi)}{(E_{0} +2\xi)^2}d\xi =
\int \frac{n(\xi)}{ E_{0}+2 \xi} d\xi
\label{de} 
\end{equation}
 We make the following assumptions about the moments (averages) of $n(\xi)$:
$<n(\xi)>  =  0$ and 
$<n(\xi) n(\xi ')>  =  n_{0}^2 \exp(-(\xi-\xi ')^2/\Delta_{1}^2)$.
This latter condition is only intended to express the fact that  the 
correlation persists
over an energy of the order of the single-particle level spacing. Taking 
the square of (\ref{de}) and averaging we obtain

$$
<\delta E^2>  (\frac{N_F}{2E_{0}})^2 = n_{0}^2 \int d\xi \int d\xi' 
\frac{e^{-(\xi-\xi ')^2/\Delta_{1}^2}}
{(E_{0} +2\xi) (E_{0} +2\xi')}.
$$
This gives
$\delta E \sim \frac{n_{0}}{N_{F}} \sqrt{\Delta_{1} E_{0}}$. 
The quantity $n_{0}/N_{F}$ should not be larger than unity.

\end{document}